\documentclass[longauth]{aa}  
\usepackage{graphicx}
\usepackage[varg]{txfonts}
\usepackage[switch, displaymath, mathlines]{lineno}
\usepackage{url,hyperref}
\usepackage{footmisc}
\usepackage{xspace}
\newcommand{\arcsource}{HESS$~$J1746$-$285}
\newcommand{\GCsource}{HESS$~$J1745$-$290}
\newcommand{\GPWN}{G0.9$+$0.1}
\newcommand{\GPWNHESS}{HESS$~$J1747$-$281}
\newcommand{\arcPWN}{G0.13$-$0.11}
\newcommand{\sgra}{Sgr\,A$^\star$}
\newcommand{\hess}{H.E.S.S.}
\newcommand{\chandra}{\emph{Chandra}}
\newcommand{\grays}{$\gamma$-rays}
\newcommand{\gray}{$\gamma$-ray}
\newcommand{\fermi}{{\emph{Fermi}-LAT}}
\newcommand{\papI}{(paper~I)\xspace}
\newcommand{\papII}{(paper~II)\xspace}
\newcommand{\papIt}{paper~I\xspace}
\newcommand{\papIIt}{paper~II\xspace}

\begin{document}

   \title{Characterising the VHE diffuse emission in the central 200 parsecs of our
     Galaxy with \hess}

   \authorrunning{\hess\ Collaboration}

   \makeatletter
\renewcommand*{\@fnsymbol}[1]{\ifcase#1\or*\or$\dagger$\or$\ddagger$\or**\or$\dagger\dagger$\or$\ddagger\ddagger$\fi}
\makeatother

\author{\tiny H.E.S.S. Collaboration
%\and H.~Abdalla\protect\footnotemark[1] \inst{1} % main scientific contributor
\and H.~Abdalla \inst{1}
\and A.~Abramowski \inst{2}
\and F.~Aharonian \inst{3,4,5}
\and F.~Ait~Benkhali \inst{3}
\and A.G.~Akhperjanian\protect\footnotemark[2] \inst{6,5} % Deceased author
\and T.~Andersson \inst{10}
\and E.O.~Ang\"uner \inst{21}
\and M.~Arakawa \inst{43}
\and M.~Arrieta \inst{15}
% ...
}

\institute{
Centre for Space Research, North-West University, Potchefstroom 2520, South Africa \and 
Universit\"at Hamburg, Institut f\"ur Experimentalphysik, Luruper Chaussee 149, D 22761 Hamburg, Germany \and 
Max-Planck-Institut f\"ur Kernphysik, P.O. Box 103980, D 69029 Heidelberg, Germany \and 
Dublin Institute for Advanced Studies, 31 Fitzwilliam Place, Dublin 2, Ireland \and 
% 5
National Academy of Sciences of the Republic of Armenia,  Marshall Baghramian Avenue, 24, 0019 Yerevan, Republic of Armenia  \and
Yerevan Physics Institute, 2 Alikhanian Brothers St., 375036 Yerevan, Armenia \and
Institut f\"ur Physik, Humboldt-Universit\"at zu Berlin, Newtonstr. 15, D 12489 Berlin, Germany \and
University of Namibia, Department of Physics, Private Bag 13301, Windhoek, Namibia \and
GRAPPA, Anton Pannekoek Institute for Astronomy, University of Amsterdam,  Science Park 904, 1098 XH Amsterdam, The Netherlands \and
% 10
Department of Physics and Electrical Engineering, Linnaeus University,  351 95 V\"axj\"o, Sweden \and
Institut f\"ur Theoretische Physik, Lehrstuhl IV: Weltraum und Astrophysik, Ruhr-Universit\"at Bochum, D 44780 Bochum, Germany \and
GRAPPA, Anton Pannekoek Institute for Astronomy and Institute of High-Energy Physics, University of Amsterdam,  Science Park 904, 1098 XH Amsterdam, The Netherlands \and
Institut f\"ur Astro- und Teilchenphysik, Leopold-Franzens-Universit\"at Innsbruck, A-6020 Innsbruck, Austria \and
School of Physical Sciences, University of Adelaide, Adelaide 5005, Australia \and
% 15
LUTH, Observatoire de Paris, PSL Research University, CNRS, Universit\'e Paris Diderot, 5 Place Jules Janssen, 92190 Meudon, France \and
Sorbonne Universit\'es, UPMC Universit\'e Paris 06, Universit\'e Paris Diderot, Sorbonne Paris Cit\'e, CNRS, Laboratoire de Physique Nucl\'eaire et de Hautes Energies (LPNHE), 4 place Jussieu, F-75252, Paris Cedex 5, France \and
Laboratoire Univers et Particules de Montpellier, Universit\'e Montpellier, CNRS/IN2P3,  CC 72, Place Eug\`ene Bataillon, F-34095 Montpellier Cedex 5, France \and
DSM/Irfu, CEA Saclay, F-91191 Gif-Sur-Yvette Cedex, France \and
Astronomical Observatory, The University of Warsaw, Al. Ujazdowskie 4, 00-478 Warsaw, Poland \and
% 20
Aix Marseille Universit\'e, CNRS/IN2P3, CPPM UMR 7346,  13288 Marseille, France \and
Instytut Fizyki J\c{a}drowej PAN, ul. Radzikowskiego 152, 31-342 Krak{\'o}w, Poland \and
Funded by EU FP7 Marie Curie, grant agreement No. PIEF-GA-2012-332350,  \and
School of Physics, University of the Witwatersrand, 1 Jan Smuts Avenue, Braamfontein, Johannesburg, 2050 South Africa \and
Laboratoire d'Annecy-le-Vieux de Physique des Particules, Universit\'{e} Savoie Mont-Blanc, CNRS/IN2P3, F-74941 Annecy-le-Vieux, France \and
% 25
Landessternwarte, Universit\"at Heidelberg, K\"onigstuhl, D 69117 Heidelberg, Germany \and
Universit\'e Bordeaux, CNRS/IN2P3, Centre d'\'Etudes Nucl\'eaires de Bordeaux Gradignan, 33175 Gradignan, France \and
Oskar Klein Centre, Department of Physics, Stockholm University, Albanova University Center, SE-10691 Stockholm, Sweden \and
Wallenberg Academy Fellow,  \and
Institut f\"ur Astronomie und Astrophysik, Universit\"at T\"ubingen, Sand 1, D 72076 T\"ubingen, Germany \and
% 30
Laboratoire Leprince-Ringuet, Ecole Polytechnique, CNRS/IN2P3, F-91128 Palaiseau, France \and
APC, AstroParticule et Cosmologie, Universit\'{e} Paris Diderot, CNRS/IN2P3, CEA/Irfu, Observatoire de Paris, Sorbonne Paris Cit\'{e}, 10, rue Alice Domon et L\'{e}onie Duquet, 75205 Paris Cedex 13, France \and
Univ. Grenoble Alpes, IPAG,  F-38000 Grenoble, France \protect\\ CNRS, IPAG, F-38000 Grenoble, France \and
Department of Physics and Astronomy, The University of Leicester, University Road, Leicester, LE1 7RH, United Kingdom \and
Nicolaus Copernicus Astronomical Center, Polish Academy of Sciences, ul. Bartycka 18, 00-716 Warsaw, Poland \and
% 35
Institut f\"ur Physik und Astronomie, Universit\"at Potsdam,  Karl-Liebknecht-Strasse 24/25, D 14476 Potsdam, Germany \and
Friedrich-Alexander-Universit\"at Erlangen-N\"urnberg, Erlangen Centre for Astroparticle Physics, Erwin-Rommel-Str. 1, D 91058 Erlangen, Germany \and
DESY, D-15738 Zeuthen, Germany \and
Obserwatorium Astronomiczne, Uniwersytet Jagiello{\'n}ski, ul. Orla 171, 30-244 Krak{\'o}w, Poland \and
Centre for Astronomy, Faculty of Physics, Astronomy and Informatics, Nicolaus Copernicus University,  Grudziadzka 5, 87-100 Torun, Poland \and
% 40
Department of Physics, University of the Free State,  PO Box 339, Bloemfontein 9300, South Africa \and
Heisenberg Fellow (DFG), ITA Universit\"at Heidelberg, Germany  \and
GRAPPA, Institute of High-Energy Physics, University of Amsterdam,  Science Park 904, 1098 XH Amsterdam, The Netherlands \and
Department of Physics, Rikkyo University, 3-34-1 Nishi-Ikebukuro, Toshima-ku, Tokyo 171-8501, Japan
% ...
}

\offprints{H.E.S.S.~collaboration,
\protect\\\email{\href{mailto:contact.hess@hess-experiment.eu}{contact.hess@hess-experiment.eu}};
\protect\\\protect\footnotemark[1] Corresponding authors
\protect\\\protect\footnotemark[2] Deceased
}

   \date{Received March 2017; accepted June 2017}

% \abstract{}{}{}{}{} 
% 5 {} token are mandatory

	\abstract{ The diffuse very high$-$energy (VHE, $> 100$\,GeV) 
	\gray\ emission observed in the central 200\,pc of the Milky     
	Way by \hess\ was found to follow the dense matter distribution 
	in the Central Molecular Zone (CMZ) up to a longitudinal distance of about 130 pc
        to the Galactic Centre (GC), where the flux rapidly decreases.
        This was initially interpreted as the result of a burst$-$like injection of 
	energetic particles $10^4$ years ago, but a recent more sensitive \hess\ analysis  
	revealed that the cosmic$-$ray (CR) density profile drops with the distance to the centre, 
	making data compatible with a steady cosmic PeVatron at the GC. 
  	In this paper, we extend this analysis to obtain for the first time a detailed 
    characterisation of the correlation with matter and to search for additional 
    features and individual \gray\ sources in the inner 200\,pc. Taking advantage 
    of 250 hours of \hess\ data and improved analysis techniques 
    we perform a detailed morphology study of the diffuse VHE emission observed 
    from the GC ridge and reconstruct its total spectrum. To test the various 
    contributions to the total \gray\ emission, we use an iterative 2D maximum 
    likelihood approach that allows us to build a phenomenological model of the 
	emission by summing a number of different spatial components. 
  	We show that the emission correlated with dense matter covers the full 
	CMZ and that its flux is about half the total diffuse emission flux.
        We also detect some emission at higher latitude likely 
	produced by hadronic collisions of CRs in less dense regions of the GC 
	interstellar medium. We detect an additional emission component 
	centred on the GC and extending over about 15\,pc that is consistent with 
        the existence of a strong CR density gradient and confirms the presence of a CR 
	accelerator at the very centre of our Galaxy. We show that the spectrum 
	of the full ridge diffuse emission is compatible with the one previously derived 
	from the central regions, suggesting that a single population of particles 
	fills the entire CMZ. Finally, we report the discovery of a VHE \gray\ source 
	near the GC radio arc and argue that it is produced by the pulsar wind 
	nebula candidate \arcPWN.}

 
  \maketitle

  %% Redefine numeric symbols for footnotes
  \makeatletter
  \renewcommand*{\@fnsymbol}[1]{\ifcase#1\@arabic{#1}\fi}
  \makeatother

%________________________________________________________________

\section{Introduction}

There is growing evidence that the Galactic Centre (GC) plays a key role in particle acceleration
in the Galaxy. The most prominent evidence is the presence of the huge Fermi bubbles extending 10\,kpc above 
and below the Galactic disk~\citep{2003ApJ...582..246B,2010ApJ...724.1044S,2014ApJ...793...64A}. 
Even if their physical origin is still unknown, the object or processes powering
them has to be within the central hundreds of parsecs around the GC. 

The \hess\ discovery of diffuse very high$-$energy (VHE, $> 100$\,GeV) \gray\ emission in the GC region extending
over 1$^\circ$ in longitude  and spatially correlating with the dense gas
of the central molecular zone (CMZ) revealed that the region (the so called GC ridge)
is pervaded by VHE cosmic rays (CRs) with a density
up to a factor 9 times larger than the value measured locally at Earth~\citep[\papI]{2006Natur.439..695A}.
The observation of this diffuse emission has also  been reported above the TeV by
the VERITAS and MAGIC collaborations~\citep{2016arXiv160208522A,2016arXiv161107095A}.
As the diffuse \gray\ emission does not follow the gas distribution
beyond 1.0$-$1.3$^\circ$, or $140-180$\,pc at 8 kpc distance, it was first interpreted as the result of
a massive impulsive injection of CRs by a source close to the GC.%~\papI.%\citep{2006Natur.439..695A}.

This picture has now dramatically changed with the recent observation of a pronounced gradient in the CR density
profile deduced from the diffuse VHE \gray\ emission in the central 200\,pc~\citep[\papII]{GCPeVatron}. 
This profile, peaking in the inner tens of parsecs,
is found to be consistent with the propagation of particles injected by a steady source located at the GC itself.
Furthermore, no cutoff is found in the \gray\ spectrum of the diffuse emission extracted
from a region at 0.15$-$0.45$^\circ$ (20$-$63 pc)  distance to the GC, making this source
the first Galactic PeVatron, which accelerates charged particles to energies of at least 1\,PeV ($10^{15}$\,eV).

In this work, we extend the analysis presented in~\papIIt 
 to further the morphology and energy spectrum studies of the VHE
 \gray\ emission in the central 200\,pc and look for additional components and sources. To do so, we take advantage
of the large dataset accumulated by \hess\ from 2004 to 2012 and of improved analysis
methods now available~\citep{2011APh....34..858B,MVAICRC2015} to perform a 2D iterative maximum likelihood analysis and extract the various spatial components
required to model the total emission from the GC ridge, an approach similar to the one used for the \hess\
Galactic plane survey~\citep[HGPS,][]{HGPS}. Since there is no reliable 3D model of the gas distribution in the CMZ, we do not
model the ridge emission with a physically motivated CR density, but resort to an empirical model made of several
components and study the residual emission to search for new individual sources or a diffuse excess of \grays.

After a presentation of the dataset and the data reduction technique employed, we detail
the 2D maximum likelihood fitting technique. We then present the result of the iterative procedure. 
We describe the various components required to reproduce the 2D morphology of the
GC ridge diffuse emission and extract the total spectrum. We show that the emission fills the entire CMZ, 
correlates with dense matter, extends at larger latitude and exhibits a spectrum compatible with the one 
previously derived in the very central part of the CMZ, revealing that PeV CRs prevade an even 
wider area in the CMZ than previously published. We also detect an extended central 
component around Sgr A$^\star$ and show it originates most likely from an enhancement of CRs
in the inner tens of pc around the GC confirming results obtained in~\papIIt. %\citet{GCPeVatron}.  
Finally, we report the detection of a new point$-$like \gray\ source \arcsource\ in the vicinity of the GC radio arc~\citep{1984Natur.310..557Y}.
After presenting its physical characteristics, we compare it to known objects in the region
and show that its position is consistent with the Pulsar Wind Nebula (PWN) candidate \arcPWN\ \citep{2002ApJ...581.1148W}.

\section{Analysis}
\subsection{\hess\ observations and data analysis}

The High Energy Stereoscopic System (\hess) is an array of imaging
atmospheric Cherenkov telescopes located at 1835\,m altitude in the
Khomas highlands of Namibia. The \hess\ array is designed to detect
and image the brief optical Cherenkov flash emitted from air showers,
induced by the interaction of VHE \grays with the Earth's
atmosphere. In the first phase of \hess, during which the data used
here were recorded, the array consisted of four 13\,m telescopes
placed on a square of 120\,m side length. \gray events were
recorded when at least two telescopes in the array were triggered in
coincidence~\citep{FunkTriggerPaper}, allowing for a stereoscopic
reconstruction of \gray events~\citep[for further details
see][]{HessCrab}.

In 2012, \hess\ entered its second phase with the addition of a
fifth, large 28\,m telescope at the centre of the array. The addition
of this telescope, which is able to trigger both independently and in
concert with the rest of the array, increases the energy coverage of
the array to lower energies. The work presented in the following
sections does not use data recorded with the large telescope. 

The data set selected for the analysis includes observations within $5^{\circ}$ of the GC position,
conducted between 2004 and 2012.  Applying standard data quality selection criteria and keeping only four$-$telescope observations 
we use a total livetime of $259$\,hours with a mean zenith angle of $22^{\circ}$.
The data are analysed with an advanced multivariate analysis procedure developed
to improve the sensitivity to weak signals and optimised for morphological studies of very extended sources~\citep{2011APh....34..858B,MVAICRC2015} . 
This analysis provides a good angular resolution of $0.077^{\circ}$ ($68\%$ containment
radius of the point$-$spread function, PSF), crucial for the study shown here, and an average energy threshold of $\sim 350$\,GeV.
\gray\ measurements by ground$-$based Cherenkov telescopes suffer from irreducible backgrounds from air showers induced by charged CRs, electrons, protons, and heavier nuclei,
which can mimic \gray\ air showers.
Modelling and subtraction techniques are needed for these CR backgrounds.
To generate such 2D background maps for the production of \gray\ excess images we use an adaptative
ring background method (suitable for crowded regions) excluding known regions of \gray\ emission from the background ring~\citep{2007A&A...466.1219B,HGPS}. 
All the results shown here are cross-checked with an alternative \hess\ calibration and analysis chain
that was used for the HGPS production \citep{HGPS}.
The instrument response functions, PSF, and \gray\ exposure are produced
following the approach described in \citet{HGPS}.
To compare the molecular matter and the \gray\ emission in the central 200\,pc
of the Galaxy to each other, a map is produced over a region that covers the full CMZ extension from $358.5^{\circ}$ 
to $1.8^{\circ}$ in longitude and $\pm 1.5^{\circ}$ in latitude. A  $0.5^{\circ} \times 0.5^{\circ}$ region centred on
HESS$~$J1745$-$303~\citep{2008A&A...483..509A} is excluded from the maps in order not to bias the fit with unrelated emission.
For the spectral analysis, the background is estimated using the reflected region method~\citep{2007A&A...466.1219B}, where 
the background is derived from circular off$-$source regions with the same angular size and \gray\ detection efficiency as 
the on$-$source region. The differential VHE \gray\ spectrum is then fitted with a spectral model by
forward folding it with the instrument response functions~\citep{2001A&A...374..895P,Jouvin15}.

%%%%%%%%%%%%%%%%%%%%%%%%%%%%%%%%%%%%%%%%%%%%%%%%%%%%%%%%%%%%%%%%%%%%%%%%%%%%%%%%
\subsection{General morphology of the GC ridge emission}
\begin{figure*}
\begin{center}
\includegraphics[width=0.8\linewidth]{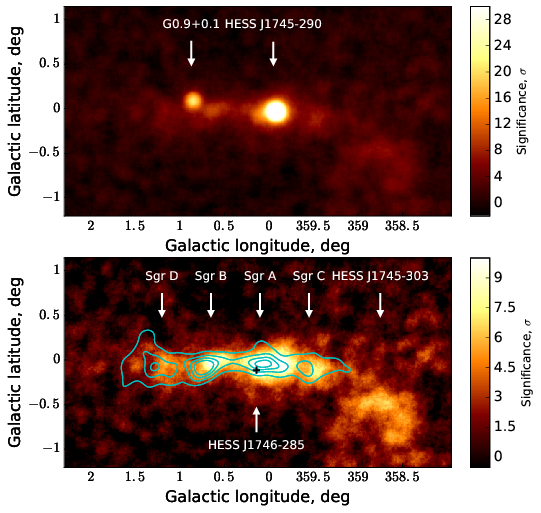}
\caption{VHE \gray\ images of the GC region in Galactic coordinates and smoothed with the \hess\ PSF.
  \textit{Top}: \gray\ significance map. \textit{Bottom}: residual significance map after subtraction of the two point
  sources G0.9$+$0.1 and \GCsource. The cyan contours indicate the density of molecular gas as traced by the CS brightness
  temperature integrated over the Local Standard of Rest (LSR) velocity from $-$200 to 200 km s$^{-1}$~\citep{1999ApJS..120....1T} and smoothed with
  the \hess\ PSF (0.077$^\circ$).  The outer contour level is 36 K km s$^{-1}$, about six times the noise level~\citep{1999ApJS..120....1T}.
  The position of the new \hess\ source \arcsource, coincident with the GC radio arc~\citep{1984Natur.310..557Y},
  is marked with a black cross.
}
\label{fig:sign}
\end{center}
\end{figure*}

Figure~\ref{fig:sign} (top) shows the \gray\ significance map of the GC region smoothed with the \hess\ PSF.
Beyond the two bright sources \GCsource\ and \GPWN, the fainter diffuse emission is visible
as well as the extended source HESS$~$J1745$-$303. 
Using the fluxes derived from the model fit (see Sect.~\ref{sec:result} below), we can produce
a significance map of the ridge emission with the point sources subtracted. 
Figure~\ref{fig:sign} (bottom) shows this map with cyan contours overlaid indicating the molecular gas density distribution as traced by the line 
of the 1$-$0 transition in the CS molecule~\citep{1999ApJS..120....1T}. 
The GC ridge emission is detected up to $\ell=1.5^{\circ}$ along the Galactic plane, following the full CMZ extension,
but with a fading brightness in its tail regions confirming the result of \papIt. %\citet{2006Natur.439..695A}.
Additionally, some faint emission at high latitude ($|b| > 0.2^{\circ}$) beyond the CS emission region is now visible.
The \gray\ map also shows more details and features about the massive molecular complexes Sgr~B2, Sgr~C,
and Sgr~D at $\ell=1.2^{\circ}$ which are clearly resolved.

%%% Figure 1  : image of diffuse emission
%%% End Figure 1

\begin{figure}
\begin{center}
 \includegraphics[width=0.5\textwidth]{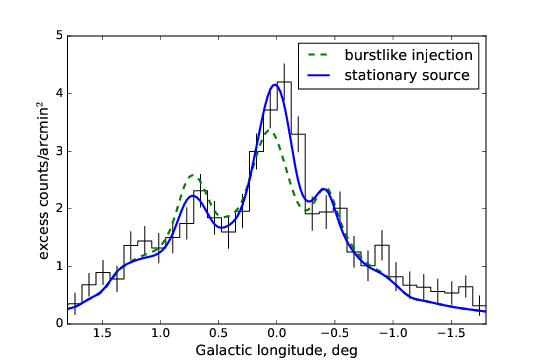}
  \caption{Longitude profile of the VHE \gray\ GC ridge emission in units of excess counts$/$arcmin$^2$.
    The CR background and the two point sources \GCsource\ and \GPWN\ have been subtracted from the counts map. The 
    profile has been integrated over $-0.3^{\circ} < b < +0.3^{\circ}$. The green line shows the model based on the VHE emission 
    profile used in \papIt, %\citet{2006Natur.439..695A},
    which includes a central $0.8^\circ$ width Gaussian multiplied by the CS map. It does not
    account for the clear excess at the GC position. As shown in \papIIt %\citet{GCPeVatron}
    the latter is rather well reproduced by
    a profile obtained with a 1/r CR density integrated over a homogeneous gas density in the line of sight and multiplied
    by the integrated CS map (blue line). %Because of the unknown 3D gas distribution, the position of the peak is slightly offset.
  }
\label{fig:profile_1overR}
\end{center}
\end{figure}

The longitude profile of the GC ridge diffuse emission is shown in Fig.~\ref{fig:profile_1overR}. 
We compare the data with the model proposed in \papIt %\citet{2006Natur.439..695A}
 to account for an impulsive injection of CRs
based on the CS map multiplied by a Gaussian of 0.8$^\circ$ width and centred at the GC.
This model fails to reproduce the central excess of emission that we clearly see in the \gray\ profile. This requires 
the presence of a strong gradient in the CR distribution peaking around the GC.
Conversely, this measured central excess of emission is well reproduced by a model where the Gaussian is replaced by 
the integrated density profile of a steady CR source~\papII. %\citep{GCPeVatron}. 
The position of the \gray\ data peak appears slightly shifted from the CR model, mainly because the emission in the central 30\,pc
is not correlated with the 2D CS distribution. Therefore, in the following, we do not try to reproduce the diffuse \gray\ emission by such
a physically motivated model that would in any case be based on the simplistic hypothesis of a homogeneous gas density along the line of sight. 
We would be forced to use such an assumption because we lack a full 3D gas distribution (see discussion in section 2.3). Since this approach does not reproduce the data well, 
we resort to building an empirical model made of multiple components that individually have no immediate physical meaning but taken together 
do reproduce the data.

%%% Figure 2 : longitude profile and 1/r CR gradient
%%% end Figure 2

%%%%%%%%%%%%%%%%%%%%%%%%%%%%%%%%%%%%%%%%%%%%%%%%%%%%%%%%%%%%%%%%%%%%%%%%%%%%%%%%
% \subsection{2D template based maximum likelihood fit} 
\subsection{Emission model fit} 
To derive the various contributions to the total \gray\ emission, we use a 2D maximum likelihood approach 
similar to the one used for the construction of the HGPS catalogue~\citep{HGPS}.
A 2D model of the expected event counts is fitted to the data assuming the events obey Poisson statistics.
The model takes into account the estimated CR backgrounds as well as a physical \gray\ model
weighted by the \gray\ exposure and convolved with the \hess\ PSF.
The different model components are:
\begin{itemize}
\item The estimated charged particle background event map ~\citep[the normalised OFF map, see ][]{HGPS}, 
left constant during the fit procedure.
\item A 2D model of the Galactic large scale unresolved emission~\citep{PhysRevD.90.122007}, left constant during the fit procedure.  
In order to estimate this contribution, we exclude regions with significant emission from the map and fit this Galactic component 
over a large box of $10^{\circ}\times 8^{\circ}$, assuming a simple emission profile with a flat distribution in Galactic longitude
and a Gaussian distribution in Galactic latitude, weighted by the \gray\ exposure~\citep[see the details of the approach in][]{HGPS}.
\item The point sources \GCsource, coincident with \sgra, and \GPWNHESS\, coincident with the composite supernova remnant (SNR) G0.9$+$0.1.
  After a first fit, the positions of these point sources are fixed but their amplitude is left free and fitted in each step of the procedure.
\item The CR induced \gray\ emission in the GC ridge. This component is assumed to be the product of a
  gas map and a 2D symmetrical Gaussian following the procedure used in \papIt %\citet{2006Natur.439..695A}
   and refered as Dense Gas Component (DGC) thereafter.
  The normalisation and Gaussian extension of this component are fitted.
\item Other ad-hoc Gaussian components are added to the model, depending on the residual map, at each step of the process. 
\end{itemize}    

The choice of the gas tracer to model the main GC ridge diffuse emission component might affect the fit result.
The  main  constituent of  the  interstellar  gas in the central 200 pc is the molecular matter distributed into a set of
very dense clouds, whereas no more than 10 percent is present in the form of atomic gas \citep{2007A&A...467..611F}.
A full 3D model of the CMZ molecular clouds would be well suited to determine the exact distribution of \gray\ emission over the ridge, 
but so far no satisfactory model is available. Indeed, our poor knowledge of the true gas kinematics near the GC renders 
the methods based on kinetics convertions between velocties and distance unreliable. 
Several 2D matter templates derived from millimeter molecular tracers of the interstellar gas distribution in the CMZ are 
available. The lines of HCN(1$-$0) and CS(1$-$0) are optically thin tracers of dense gas and are bright enough to be detected 
over most of the CMZ, whereas other molecules are detected only in the densest cores~\citep{2012MNRAS.419.2961J}.
Here, we use the CS (1$-$0) line emission~\citep{1999ApJS..120....1T}.   
To test the impact of the actual gas tracer used, we tested the HCN (1$-$0) transition from \citet{2012MNRAS.419.2961J} and 
obtained consistent results confirming that CS is a robust tracer. We will therefore only show results obtained with CS in the following. 
We note that all these surveys are only sensitive to dense gas ($n > 10^4$ cm$^{-3}$). We therefore expect our main ridge component
based on CS to miss regions of more diffuse gas. 
Furthermore, most tracers suffer from self$-$absorption in extremely dense
regions like the core of Sgr\,B2. 
The model must be poor in this specific region. 
The recent high-angular resolution column density map of the central part of the CMZ derived from thermal images
of cold dust \citep{2011ApJ...735L..33M} does not suffer from this problem but its sensitivity
to the high ionization rate in the core of Sgr\,B2 can also affect the derived density.

The iterative fitting process is performed using the Sherpa software package~\citep[CIAO v4.5, ][]{Sherpa}. The model parameter values
and associated confidence intervals are given for the best fit model in Table~\ref{table:fit_results}, and each step is detailed in
Table~\ref{table:2} in the appendix. 
The model is tested against \hess\ data during the fitting procedure and two criteria are guiding the search for
the best model: a good statistical fit and the distribution of residuals. The significance of different components is measured by the 
likelihood ratio test statistic (TS), $\rm TS= 2 \rm  log ( \rm  L_{n+1}/ \rm L_{n}$), comparing the likelihood between model $n$ and model $n+1$. 
At each step, a significance map of the residual and a TS map is created, allowing to search for features 
that should be included in the model. 
This comparison of TS values remains an important tool for selecting models that provide gradual improvements when fitting the data.
In the limit of a large number of counts per bin the TS for the null hypothesis is asymptotically distributed as a $\chi^2 / d.o.f.$, 
where $d.o.f.$, the degrees of freedom, is the number of free parameters describing the different model components. The detection 
threshold is set to TS $>30$ following the prescription used for the HGPS \citep[see Sect. 4.8 of][]{HGPS}.\\
 
In order to quantify the fit quality, we use the Sherpa implementation of the \textit{Cash} statistic~\citep{1979ApJ...228..939C},
CSTAT\footnote{http://cxc.harvard.edu/sherpa4.4/statistics/\#cstat}, divided by the number of degrees of freedom, which is of the 
order of 1 for good fits. Since this criterion may only be used if the number of counts in each bin is high, we rebin the maps to 
ensure this condition is fulfilled. We also check that the statistical significances of the residuals at each position are consistent
with those expected from a large number of Monte Carlo simulations assuming the final model.

%%% Figure 3 : iterative fitting
\begin{figure*}
\centering
\includegraphics[width=16cm,clip]{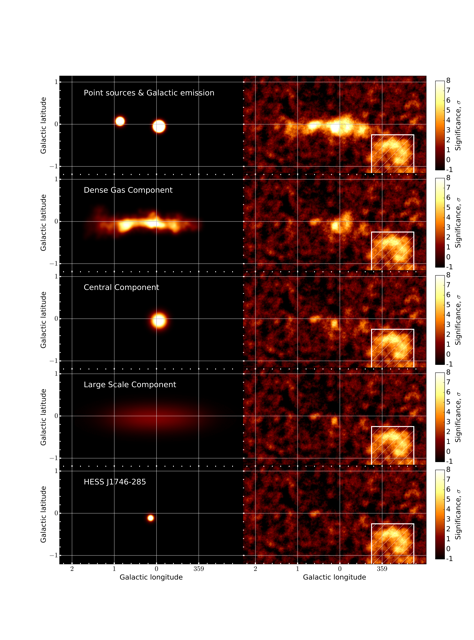}
\caption{
The successiv stages of the iterative fitting process are shown here.
The model count map of each individual component (correlated with the \hess\ PSF) added in each step of the analysis is shown in the left$-$hand column 
(in units of counts per pixel), the corresponding residual significance map (in units of significance level) in the right$-$hand column, both in Galactic coordinates. 
From top to bottom we show: background $+$ \GPWN\ $+$ \GCsource\ $+$ Galactic large scale unresolved, the dense gas component (DGC)
, the central component (CC), the large scale component (LSC) and the new source \arcsource.
The residual Li \& Ma significance maps are computed using the data counts map as signal and the model map (correlated with the \hess\ PSF) as background.}
\label{appfig}
\end{figure*}
%%%% end figure 3

%% Figure 4: longitude and latitude profiles. Comparison with models.
\begin{figure*}
\begin{center}
  \includegraphics[width=\textwidth]{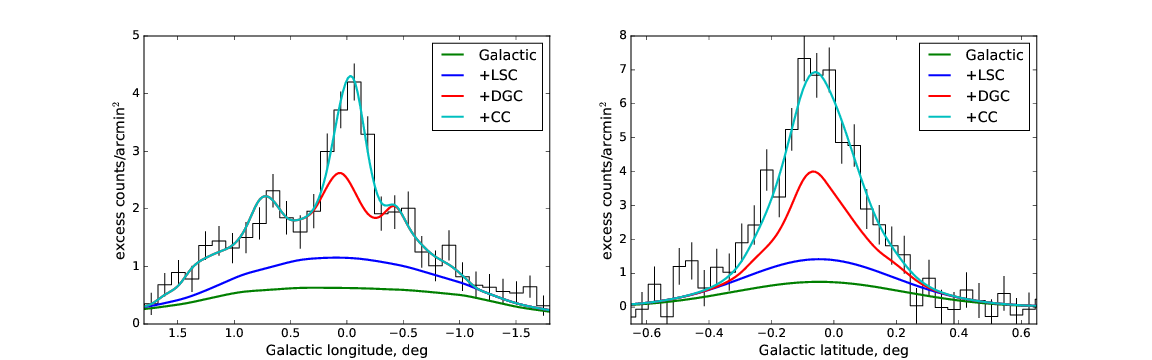}
  \caption{New longitude \textit{(left)} and latitude \textit{(right)} profiles of the VHE \gray\ GC ridge emission in unit
    of excess counts$/$arcmin$^2$.
    The normalised background map and the two point sources \GCsource\ and \GPWN\ have been subtracted from the \gray\ counts map using
    the fitted model parameters. The longitude profile has been integrated over $-0.3^{\circ} < b < +0.3^{\circ}$ and the latitude
    profile over $359.5^{\circ} < l < + 0.5^{\circ}$. Coloured lines show how the different components of the model reproduce the data.
    In particular, it can be seen that the large scale component (LSC) accounts for the high latitude residual emission, and
    the central component (CC) accounts for the central residual excess. The Galactic Large Scale emission \citep{PhysRevD.90.122007}
    does not contribute signficantly in this region.}
\label{fig:profiles}
\end{center}
\end{figure*}
%%% end figure 4

%%%%%%%%%%%%%%%%%%%%%%%%%%%%%%%%%%%%%%%%%%%%%%%%%%%%%%%%%%%%%%%%%%%%%%%%%%%%%%%%
%\section{Results}  \label{sec:result}
\subsection{Results of the emission model}  \label{sec:result}

%%%%% Table 1: final fit parameters
\begin{table*}
\caption{Parameters describing the components for the final model.
  The background normalisation and the Galactic large scale emission are unchanged by the fit and determined beforehand as explained in the main text.
  The two sources \GPWN\ and \GCsource\ are modelled as point sources. The three components modelling the GC ridge emission are the dense gas component %(CS\,$\times$\,CR)
  (DGC), the large scale component (LSC), and the additional central component (CC).
  The parameter values and associated statistical errors at $68\%$ confidence$-$level are indicated, as well as a systematic uncertainty
  obtained by combining a 30\% flux normalisation uncertainty~\citep{HGPS} and a background uncertainty (obtained after performing the iterative fitting
  procedure including a variation of the background component level by $\pm$ 2\%).} 
  \centering
\begin{tabular*}{\textwidth}{@{\extracolsep{\fill}} c c c c}      
\hline\hline\noalign{\vskip 2mm}              
 Model        &  Position                & Extension    &     Flux                                                    \\
 component   & (Galactic Coordinates)   & (degrees)    &    $ ( 10^{-12}  \rm cm^{-2} \rm s^{-1} \rm  TeV^{-1} )$  \\\noalign{\vskip 2mm}\hline\hline\noalign{\vskip 2mm}
        & $\ell= 0.86^{\circ}$        &              &                     \\
\raisebox{1.5ex}[-1.5ex]{\GPWN}         & $b=0.069^{\circ} $        &  \raisebox{1.5ex}[-1.5ex]{--} &  \raisebox{1.5ex}[-1.5ex]{$0.88  \pm 0.04_{stat}  \pm 0.25_{sys} $} \\\noalign{\vskip 2mm}
         \hline\noalign{\vskip 2mm}
    & $\ell=  359.94^{\circ}  $   &              &                        \\
\raisebox{1.5ex}[-1.5ex]{\GCsource}              & $b= -0.05^{\circ} $      & \raisebox{1.5ex}[-1.5ex]{--} & \raisebox{1.5ex}[-1.5ex]{$2.9 \pm 0.4_{stat}   \pm 0.8_{sys} $} \\\noalign{\vskip 2mm}
\hline \noalign{\vskip 2mm}
%CS\,$\times$\,CR 
Dense Gas (DGC) &  $\ell=0^\circ$, $b=0^\circ$    & $\sigma  = 1.11^{\circ}\pm0.17^{\circ}_{stat} \pm 0.17^{\circ}_{sys}$   & $4.3  \pm 0.9_{stat}  \pm 1.5_{sys}  $\\\noalign{\vskip 2mm}
\hline\noalign{\vskip 2mm}
Central (CC)             &  $\ell=0^\circ$, $b=0^\circ$ & $\sigma  = 0.11^{\circ}\pm0.01^{\circ}_{stat}\pm 0.02^{\circ}_{sys} $     & $1.03 \pm 0.05_{stat}   \pm 0.25_{sys} $ \\\noalign{\vskip 2mm}
\hline\noalign{\vskip 2mm}
            &   &  $\sigma_{\rm x} = {{0.97^{\circ}}^{+0.04^{\circ}}_{-0.02^{\circ}}}_{stat} \pm 0.13^{\circ}_{sys}$  &  \\
\raisebox{1.5ex}[-1.5ex]{Large Scale (LSC)} & \raisebox{1.5ex}[-1.5ex]{$\ell=0^\circ$, $b=0^\circ$} &  $\sigma_{\rm y} = 0.22^{\circ}\pm0.06^{\circ}_{stat} \pm 0.07^{\circ}_{sys}$  & \raisebox{1.5ex}[-1.5ex]{$2.68\pm 0.6_{stat}  \pm 1.3_{sys}  $}\\\noalign{\vskip 2mm}
\hline\noalign{\vskip 2mm}
    & $\ell= 0.14^{\circ}\pm0.01^{\circ}_{stat} \pm 0.01^{\circ}_{sys} $ & $\sigma_{\rm x} = 0.03^{\circ} \pm 0.03^{\circ}_{stat} \pm 0.03^{\circ}_{sys} $ & \\
\raisebox{1.5ex}[-1.5ex]{\arcsource} & $b= -0.11^{\circ}\pm 0.02^{\circ}_{stat} \pm 0.02^{\circ}_{sys} $  &    $\sigma_{\rm y} = 0.02^{\circ}\pm 0.02^{\circ}_{stat} \pm 0.03^{\circ}_{sys}$   &  \raisebox{1.5ex}[-1.5ex]{$0.24 \pm 0.03_{stat}  \pm 0.07_{sys} $}\\\noalign{\vskip 2mm}
\hline\hline
\end{tabular*}
\label{table:fit_results}   
\end{table*}
%%%%%%%% End Table 1

In this section we present our main results from the template-based likelihood fit analysis.
The individual components that contribute to the final model are added step by step as illustrated
in Fig.~\ref{appfig}. The final fit result is summarized in Table~\ref{table:fit_results},
and the results of the individual components are detailed in Table~\ref{table:2}.
The contribution of the model components to the total \gray\ emission is shown in the form of
longitude and latitude profiles in Fig.~\ref{fig:profiles}.

%%%%CS
The dominant dense gas component (DGC) %(CS\,$\times$\,CR)
is modelled by the dense matter template multiplied by a Gaussian centred on 
$|\ell|=0^\circ$, $|b|=0^\circ$ with extension $\sigma$. The latter is not physically 
motivated but required by the data. It provides the characteristic scale of the extension of CRs in the GC ridge region. 
With a fitted value of $\sigma=1.1^{\circ}$ the extension is found to be slightly larger than previously estimated in 
\papIt 
, where $\sigma=0.8^{\circ}$ was found to yield the best fit to the data. 
The DGC represents a large flux fraction of  $ \sim 50\%$ of the total GC ridge emission, 
confirming that the latter is dominantly from CRs interacting with dense matter.

%%%%%%% LS
An additional large scale component (LSC), modelled as a 2D Gaussian with position fixed on SgrA$^\star$ and normalisation and extension 
($\sigma_{\rm X}$, $\sigma_{ \rm Y}$) fitted as free parameters, is required to reproduce the \gray\ emission especially at large latitudes. 
This component extends $\pm30$\,pc in latitude and $\pm 150$\,pc in longitude, and represents a non$-$negligible 
flux fraction of $30\%$ of the total GC ridge emission. Its possible origin will be discussed in section 3.

%%%% CC
An additional extended \gray\ emission excess is also detected in the very central 30\,pc region, as illustrated by the residuals 
in Fig.~\ref{appfig}. This central component (CC), modelled by a 2D symmetric Gaussian centred on SgrA$^\star$, is detected at $8.7 \sigma$
significance level. It has an intrinsic extension of $0.1^{\circ}$ and its flux represents $\sim 15\%$ of that of the ridge.
Representing a fraction of $\sim 30\%$ of the GC source \GCsource\ and an intrinsic extension of almost twice the \hess\ PSF, we can rule out 
the conclusion that this component arises from a contamination from the central point source due to systematic 
uncertainties of the PSF~\citep[see][]{HGPS}.
As we will discuss in the next section, this additional component is in good agreement with the physical interpretation
developed in \papIIt. 

%%%ArcSource
After the subtraction of the central component, a significant but localized excess is still visible in the residuals
(see fourth row of Fig.~\ref{appfig}). Adding a Gaussian component at this position results in an increase of
$\Delta$TS$=$48 or a significance of about 5.9$\sigma$. Changing the dense gas tracer as a model basis in the CMZ 
does not strongly change this significance, confirming the detection of a new source in this region.
Called \arcsource, this new source is located at Galactic position $\ell= 0.14^{\circ}\pm0.01^{\circ}, b = -0.11^{\circ}\pm0.02^{\circ}$
and has no significant extension within uncertainties. For further discussions of this new \gray\ source see Sect.~\ref{sec:arcsource} below.
After this final step in the fitting procedure, no further component is found to be significant in the residuals and the iteration is stopped. 

The major source of systematic uncertainties of our measurement is the imperfect estimate of the charged CR induced background of the \hess\ image.
In order to study how robust the inferred model components and parameter values are against systematic variations of this background, we artificially change the 
background normalisation by $\pm 2\%$, the typical systematic background uncertainty~\citep{2007A&A...466.1219B},
and reapply the full iterative procedure to rederive the model components. 
This allows the estimation of typical systematic intervals for each parameter and the testing of the robustness of a component detection. 
The LSC is the one most sensitive to background uncertainties with a resulting variation of $ \pm 50\%$ in amplitude,
around $\pm 15\%$ in longitude and $\pm 25\%$ in latitude extensions.
The uncertainty on the DGC component amplitude is also large, $\pm 30\%$. The associated spatial extension is relatively stable (within $10\%$).
The other more localized components remain stable. The systematic errors given in Table~\ref{table:fit_results} also include
a global flux normalisation uncertainty of about $30\%$ added in quadrature.

%%%%%%%%%%%%%%%%%%%%%%%%%%%%%%%%%%%%%%%%%%%%%%%%%%%%%%%%%%%%%%%%%%%%%%%%%%%%%%%%
\section{GC Ridge Diffuse Emission}
\subsection{Morphology}
The results given above confirm that most of the diffuse GC ridge \gray\ emission is distributed 
like the dense gas in this region. About 50\% of the emission is found to closely follow the CS template up to a projected longitudinal distance of 
$\sim$ 1.0$^\circ$ or 140\,pc from the GC. The dip in \gray\ emission beyond 100$-$150\,pc
is likely from a combination of decreasing CR density with distance to the GC and a more diffuse matter distribution
along the line$-$of$-$sight. This notion is supported by the face$-$on view of the CMZ provided by \citet{2004MNRAS.349.1167S}, which shows that the region around 1.3$^\circ$
is much more spread out along the line$-$of$-$sight than the central part of the CMZ.

We find that an additional large-scale emission, LSC above, is required to reproduce the observed morphology. Even if part of 
this diffuse emission is sensitive to systematic uncertainties of charged CR backgrounds, its detection is clearly significant beyond
statistical and systematic uncertainties. This component does not correlate with dense gas tracers; in particular its latitude extension is larger. 
The reason is that CS does not fully trace both the densest CMZ structures and the more diffuse gas. It has in fact been estimated that about
30\% of the molecular gas in the region is found in a diffuse H$_2$ phase of density $\sim$\,100\,cm$^{-3}$~\citep{1998A&A...331..959D}.
This is in good agreement with the relative contribution of the LSC to the total GC ridge emission. A natural explanation for it
is therefore that most of the large scale component flux is from CRs pervading the central 200\,pc and interacting with this
diffuse phase of molecular hydrogen. 
The atomic gas could also be invoked as it has been established as an important ISM component for some TeV sources. However 
it represents no more than 10 percent of the gas in the CMZ \citep{2007A&A...467..611F}, which is not enough to explain all of the LSC. 
Further studies could however be carried out using recent high resolution HI images \citep{2012ApJS..199...12M}.
 
We note that IC emission from VHE electrons in the region cannot be formally excluded, but requires a broad distribution of sources
in the inner 200 pc. The radiative losses are large enough to prevent propagation on such scales. The spectral shape of this
component should then be significantly different from the dense gas one.
Unresolved sources could also contribute to the emission.
Finally, some contribution from the densest structures in the CMZ is also likely, as suggested by the presence of a hotspot at the
position of the core of Sgr\,B2, which is however not significant enough to be detected as an individual source component.

Finally, the most interesting feature we find is the central component, CC, which is centred on the GC
and has a radius of about 15\,pc. It further proves the presence of a gradient of CRs peaking around
the GC as shown in \papIIt. %\citet{GCPeVatron}. 

To quantify this, we estimate the ratio of the energy density of
cosmic rays, w$_{\rm CR}$ (in eV cm$^{-3}$), confined in the central
30\,pc, to the average density in the whole CMZ, considering the
extensions and luminosities of the CC and DGC %(CS\,$\times$\,Gauss)
components, respectively.  We assume that the gas density $n_{\rm H}$ is
uniform in the region. The volume $V_{\rm CMZ}$ containing 50\% of the
GC ridge flux is a cylinder of 170\,pc radius and 30\,pc
thickness. The average CR density in this region is
$w_{\rm CMZ} \propto L_{\rm CMZ}/V_{\rm CMZ}$ where $ L_{\rm CMZ}$ is total VHE luminosity. 
Similarly, the 50\% flux equivalent volume of the central component is a sphere of 23\,pc
radius. Therefore the ratio of CR density in the central 20\,pc over
the density in the whole CMZ is
$\propto L_{\rm CC}/ L_{\rm CMZ} \times V_{\rm CMZ}/V_{\rm CC} \sim 10$.  If we
estimate this ratio for CRs following a 1/r profile, as expected for a
steady source~\papII, %\citep{GCPeVatron},
we indeed confirm that the energy
densities between the central region and the entire CMZ differ by a
factor on the order of 10. This shows that the empirical model used to
reproduce the GC ridge morphology is fully compatible with the stationary
source hypothesis: a large fraction, if not all, of the CRs pervading
the CMZ are accelerated at the GC or its direct vicinity.

Unresolved sources might also contribute to this central excess, in particular SNRs, given the high supernova rate in the CMZ.
On the other hand, the soft thermal X$-$ray emission that traces SNRs in the CMZ is much more
extended~\citep[up to 0.2$^\circ$$-$0.3$^\circ$  around the GC,][]{2015MNRAS.453..172P}
and it fails to reproduce the morphology of the VHE \gray\ emission measured with \hess\ Similarly,
the electrons injected by the population
of PWNe detected with \chandra\ in the inner 30\,pc~\citep{2008ApJ...673..251M,2009MNRAS.399.1429J} could potentially provide
a large \gray\ luminosity. But given the large infrared (IR) and optical photon field energy densities in the GC region,
the measured energy spectra should show pronounced Klein$-$Nishina suppression effects in the TeV \gray\ range,
which is inconsistent with the hard spectrum without any cut$-$off reported in \papIIt %\citet{GCPeVatron}
in the central 0.45$^\circ$.
The PWNe scenario is therefore highly disfavoured. The central component is very likely due to a peak of the CR density
caused by an accelerator at or very near to the GC. The overall diffuse \gray\ emission measured
from the central 200\,pc of the Milky Way is consistent with CRs injected by this central accelerator, diffusing away
and interacting along the GC ridge.

\subsection{Total Spectrum}
The total GC ridge spectrum is evaluated over the large area $\lvert \ell \lvert<1^{\circ}$, $\lvert b \lvert <0.3^{\circ}$ by dividing it into smaller regions,
performing the spectral analysis in each region, and summing up the energy spectra to perform a global fit over all regions together.
11 rectangular regions of similar area $ (2.7 \times 10^{-5} \rm sr)$ are defined over the GC ridge. 
They cover most of the diffuse emission and are chosen to have a surface brightness as similar as possible. 
Two regions of $0.2^{\circ}$ radius around G0.9$+$0.1 and Sgr\,A$^\star$ are excluded from the analysis ensuring negligible contamination ($<2\%$) from 
the bright point sources. A region of $0.1^{\circ}$ around the new source HESS$~$J1746$-$285 is also excluded
from the analysis.

%%%Figure 5   Ridge spectrum
\begin{figure}
%\begin{center}
%\includegraphics[width=0.5\textwidth]{spectrum_ridge_v2.eps}
\includegraphics[width=0.5\textwidth]{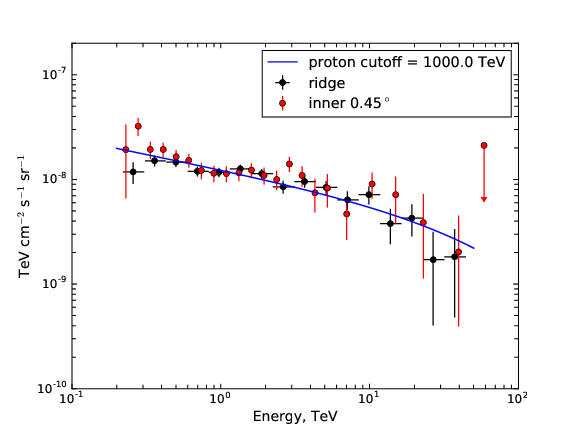}
\caption{VHE \gray\ flux per unit solid angle in the Galactic Centre region (black data points). 
  Shown is the spectrum of the GC ridge region, $|\ell|\ <\ 1^{\circ}$, $\lvert b\lvert\ <\ 0.3^{\circ}$. All error bars show the 1$\sigma$ standard deviation and 
  are corrected to account for some background double counting due to the stacking procedure. The spectrum is fitted over
  an energy range up to 45 TeV. It can be described by a power law with a photon index of
  $2.28 \pm 0.03_{ \rm stat} \pm 0.2_{ \rm syst}$ and a differential flux at 1\,TeV of 
  $1.2 \pm 0.04_{ \rm stat} \pm 0.2_{ \rm syst} \times  10^{-8} \rm TeV^{-1}  \rm cm^{-2}  \rm s^{-1}  \rm sr^{-1}$.
  For comparison, the blue line is the \gray\ spectrum resulting from a power$-$law proton spectrum with a cutoff at 1\,PeV.  
  }
\label{fig:Ridge_spectrum}
%\end{center}
\end{figure}
%%% End figure 5

Under the hypothesis that the spectral characteristics do not vary
across the ridge, all individual regions have been analysed
independently, stacked up, and fitted to obtain the average energy
spectrum shown in Fig. \ref{fig:Ridge_spectrum}. This method ensures a
small dispersion of observation conditions and instrument response
functions per region when building the individual spectra. It also
permits choosing a sufficient number of background regions, following
the reflected region method~\citep{2007A&A...466.1219B}, for each
individual rectangular region, which for one single very large region
(i.e.\ the whole GC ridge) for spectral analysis would not be possible.
Following this procedure, some of the background regions overlap with
each other leading to double counting of background events; the
resulting statistical uncertainties are thus too small by 25\% and the
uncertainties of the resulting total spectrum shown in
Fig.~\ref{fig:Ridge_spectrum} are corrected for this.

The derived spectrum of the entire GC ridge can be described by a power
law with a hard photon index of
$2.28 \pm 0.03_{ \rm stat} \pm 0.2_{ \rm syst}$.  This result is
compatible with the previous spectrum obtained in \papIt %\citet{2006Natur.439..695A}
 considering the larger analysis region
used here.  The spectrum is fitted in an energy range that extends
up to 45\,TeV, a maximum energy beyond which our systematic
uncertainty for such a faint signal does not allow to constrain the
spectral shape robustly. No significant spectral cut$-$off is found. We
compare the \gray\ spectrum with a simple power law model of the
parent proton population with high energy cutoff at 1\,PeV in
Fig.~\ref{fig:Ridge_spectrum}. As seen in this figure, the flux points
are in good agreement with a proton cutoff energy at or above 1\,PeV,
as well as with the flux points from the inner 0.45$^\circ$ (or 63 pc) 
obtained in \papIIt. %\citet{GCPeVatron}.

The total ridge \gray\ flux in the box $|\ell| <1^{\circ}$,
$|b| <0.3^{\circ}$ is estimated to
$\Phi(>1\ \rm TeV)= 1.5 \times 10^{-11}\ \rm TeV\ \rm cm^{-2}\ \rm
s^{-1}$, which corresponds to a total \gray\ luminosity of
$\rm L_{\gamma}(>1\ \rm TeV) = 2 \times 10^{35}\ \rm erg\ \rm s^{-1}$ at
the GC distance. This is slightly smaller than the value derived from
the 2D analysis but considered consistent with it in view of the
limited extension considered here. Since the \gray\ emission is
entirely due to the decay of neutral pions produced by proton$-$proton
interactions, we can express the total energy of CR protons as
$W_{\rm CR} = L_{\gamma} \times t_{pp}$, with $L_{\gamma}$ the
\gray\ luminosity and $t_{pp}=1.6 \times 10^8\ \rm yr\ (n_{\rm H}/1 \rm cm^{-3})^{-1}$,  where $n_{\rm H}$ is the
gas density. This implies a total CR energy of
$W_{pp} (>10\ \rm TeV) \sim 1 \times 10^{49}$\,erg confined in the CMZ
for a typical value of $n_{\rm H} \sim 100 \,\rm cm^{-3}$.

%%% Figure 6
\begin{figure}
%\begin{center}
%\includegraphics[width=0.5\textwidth]{arc_spec2.eps}
\includegraphics[width=0.5\textwidth]{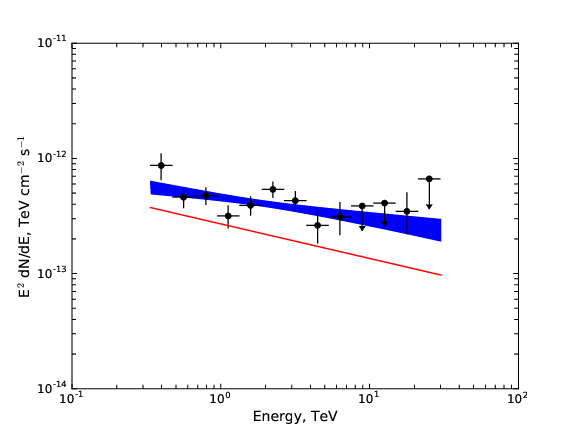}
\caption{The VHE \gray\ spectrum of the region centred on the position
  of \arcsource, fitted with the sum of two power laws. 
  The GC ridge contribution is fixed and the intrinsic source spectrum of
  \arcsource\ is fitted to the data.  In red, we show the fixed ridge
  power law contribution to the total spectrum. The intrinsic spectrum of \arcsource\ was estimated to have a flux normalisation of
  $\text{F}(1 \text{TeV})= (1.8 \pm 0.5) \times 10^{-13} \text{cm}^{-2} \text{s}^{-1} \text{TeV}^{-1}$ and an
  index of $2.2 \pm 0.2$ for the energy range above $0.350$ TeV. The errors include the uncertainty of the GC ridge emission, which
  are obtained by varying the ridge component parameters by their statistical errors. } 

  \label{fig:arcsource_spectrum}
%\end{center}
\end{figure}
%%% end figure 6

%%%%%%%%%%%%%%%%%%%%%%%%%%%%%%%%%%%%%%%%%%%%%%%%%%%%%%%%%%%%%%%%%%%%%%%%%%%%%%%%
\section{A new VHE source in the vicinity of the GC radio arc: \arcsource} \label{sec:arcsource}
\subsection{Position and spectrum}
\arcsource\ is detected at Galactic position
$\ell= 0.14^{\circ}\pm 0.01^{\circ}, b = -0.11^{\circ}\pm 0.02^{\circ}$. 
It is important to note that the systematic
uncertainties on the knowledge of the exact morphologies used for the
various components to model the GC ridge diffuse emission can be large. To
estimate these, we evaluate the impact of using HCN rather than CS as
tracer for dense gas and also the impact of changing the normalisation
of the CR background. We find the systematic error on the source
position to be $0.02^\circ$. No significant extension is found and we
place an upper limit on the extension at $0.05^\circ$.

Because of the relatively large intensity of the GC ridge emission, extracting the intrinsic
spectrum of \arcsource\ requires special care. The spectrum of the diffuse emission in the source extraction
region must be estimated and included in the modelling of the total observed spectrum.

The spectral extraction is therefore performed in two circular regions
with radii of $0.09^\circ$.  One is centred on the best-fit position
of \arcsource\ and the other on the symmetric position with respect to
the GC. The diffuse emission spectral contribution can then be evaluated on the
second region and included in the modelling of the observed counts of
the source region after renormalising the flux according to the ridge emission
morphology obtained with the 2D analysis. The implicit assumption made
here is that its spectral shape does not vary significantly over
the central 50\,pc.

After applying the flux correction factor, found to be $\sim$ 0.9, the
parameters are fixed and used to model to spectrum of the source
region, which consists of the charged CR background, the ridge emission and the
intrinsic spectrum of \arcsource\ described by a power law.  The
intrinsic spectrum of \arcsource\ has a flux normalisation of
$\text{F}(1~\text{TeV})= (1.8 \pm 0.5) \times 10^{-13} \text{cm}^{-2}
\text{s}^{-1} \text{TeV}^{-1}$ and an index of $2.2 \pm 0.2$ for the
energy range above $0.350$~TeV (see
Fig.~\ref{fig:arcsource_spectrum}).  The errors include the
uncertainty of the ridge emission, which are obtained by varying the
ridge component parameters by their statistical errors.  This gives a
luminosity in the range 0.35$-$10~TeV of about
$7 \times 10^{33} \text{erg\ s}^{-1}$ at 8 kpc distance.

Both MAGIC and VERITAS collaborations have recently found significant excesses
in the vicinity of \arcsource ~\citep{2016arXiv160208522A,2016arXiv161107095A}, but
their studies are based on much lower observation times and do not take into account the contribution
of underlying diffuse emission. The existence of an excess of diffuse emission around this position is
clearly visible from the early images of the GC ridge emission in \papIt and is due to
the presence of the massive molecular complex Sgr A. The existence of a faint source
(below the percent Crab flux) on top of this diffuse emission requires a detailed modeling of the latter
as  described above. We therefore consider VER$~$J1746$-$289 and MAGIC$~$J1746$-$285 to be a mix of
the ridge emission, which is very intense around the Sgr A molecular complex and of \arcsource\ emission. 

%%% Figure 6
\begin{figure*}
\begin{center}
\includegraphics[width=0.49\textwidth]{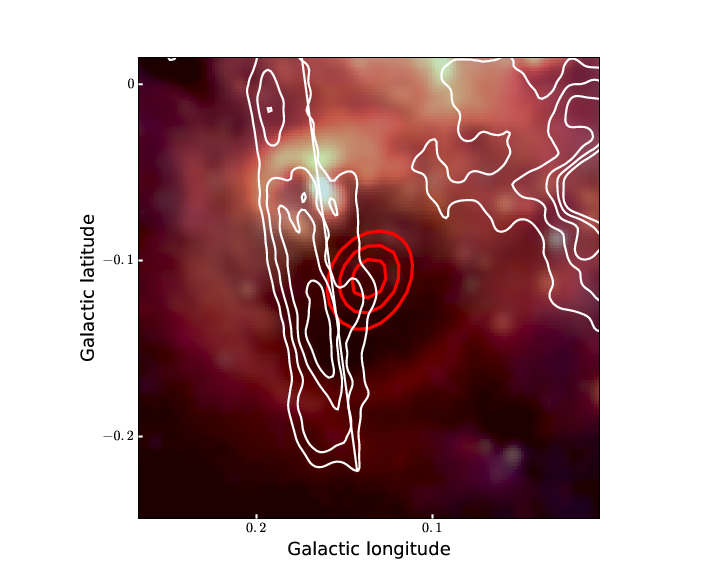}
\includegraphics[width=0.49\textwidth]{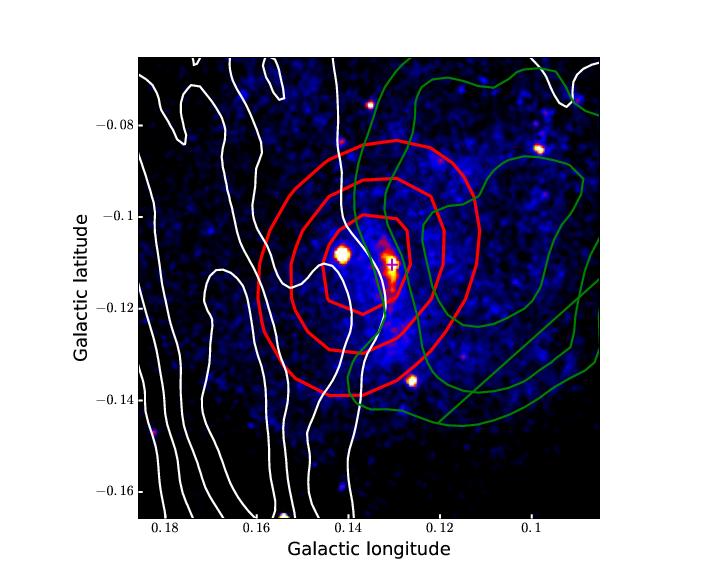}
\caption{\emph{Left:} Three colour MSX \citep{2001AJ....121.2819P} IR image of the radio arc bubble (band C, D and E).
  Red contours overlaid show the confidence$-$level
  contours at 68\%, 95\%, and 99\% on the best fit position of the new detected source \arcsource\ derived from the TS map of
  the fitted position, white contours show the radio arc emission at 90 cm. The Quintuplet cluster is the bright source on the
  top left part of the bubble.
  \emph{Right:} \chandra\ 2$-$10 keV mosaic image of the field surrounding \arcPWN\ using all available public data.
  The image is exposure$-$corrected and smoothed with a Gaussian with a width of $\sigma$ = 2.5'' to highlight the filamentary
  emission.  Green contours show the HCN emission of the molecular cloud G0.11$-$0.11 integrated in the range
  10$-$30 km/s~\citep[following][]{2011PASJ...63..763T}.}
\label{fig:arcsource_mwl}
\end{center}
\end{figure*}
%%% End figure 6

\subsection{Nature of the source}\label{sec:arc_source_nature}
%% The global picture
\arcsource\ is positionally coincident with the confused \fermi\ source 3FGL~J1746.3$-$2851c
at $\ell=0.149^{\circ}$, $b=-0.103^{\circ}$~\citep{2015ApJS..218...23A}. The spectrum of the latter is found to be significantly
curved with a log$-$parabola shape with index 2.42 $\pm$ 0.07 and $\beta = 0.34 \pm 0.06$. 
It is also present in the 1FHL catalogue as 1FHL~J1746.3$-$2851~\citep{2013ApJS..209...34A}, with a large flux
above 1~GeV of $6.6 \times 10^{-9} ~ \rm ph ~ \rm cm^{-2} ~ \rm s^{-1}$ and a very soft index of $3.2  \pm 0.3$. It
is not listed in the catalogue of sources above 50~GeV~\citep{2016ApJS..222....5A}.

%%%% The IR bubble & the quintuplet
\arcsource\ lies on the edge of the so$-$called GC radio arc~\citep{1984Natur.310..557Y} at the centre
of the radio arc bubble, a large (9\,pc diameter) open circular
structure visible in mid IR
images~\citep{1999ESASP.427..699L,2001A&A...377..631R}, as shown in
the left panel of Fig.~\ref{fig:arcsource_mwl}. 
This bubble is apparently connected to the Quintuplet cluster, which is one of the three young massive stellar clusters
in the GC region. Its estimated age is $\sim$ 4~Myr and its luminosity
is $1.2 \times 10^{41}$~erg/s~\citep{1999ApJ...514..202F}.
It is the dominant source of optical/UV radiation in the region, responsible for
most of the ionization in the bubble~\citep{2007ApJ...670.1115S}. 
The radio arc bubble is filled with soft thermal X$-$rays, likely
produced by supernova explosions and massive stellar winds from 
the Quintuplet cluster~\citep{2015MNRAS.453..172P}.

%%%% The Radio Arc

%%%% Chandra & the PWN G0.13-013 
Figure~\ref{fig:arcsource_mwl} (right panel) shows the confidence$-$level contours on the best fit
position of \arcsource\ superimposed on a 2$-$10~keV exposure$-$corrected \chandra\ image.  The source is
coincident with an X$-$ray non$-$thermal filamentary structure called
G0.13$-$0.11, which has been proposed to be a PWN candidate~\citep{2002ApJ...581.1148W}. 
The luminosity derived from \chandra\ observations for G0.13$-$0.11 in the 2$-$10 keV range is L = $3 \times 10^{33} \rm erg/s$, with a spectral
photon index between 1.4 and 2.5. \citet{2002ApJ...581.1148W}
estimated the spin$-$down power of the putative pulsar, $\dot{E}$, to be on the
order of $10^{35}$~erg/s. Given the good positional coincidence and
point$-$like nature of the VHE \gray\ source, we consider \arcPWN\ to be the most likely
origin of \arcsource. We explore this scenario further below.

% Check more recent constrains on spectral index
The X$-$ray PWN is a filament of 2''$\times$40''. The cooling time of electrons along the filament imposes an upper limit
on the magnetic field of 300~$\mu$G. The thickness of the filament also constrains the magnetic field to be larger than
$20\ \mu G\ (\dot{E}/10^{35} \text{erg/s})^{1/2}$~\citep{2002ApJ...581.1148W}. No significant cooling is visible along the filament
\citep{2009MNRAS.399.1429J}.

%%%% The radiation field
We can estimate the optical radiation field energy density produced by the nearby Quintuplet cluster at the position of the
PWN candidate assuming it lies at the projected distance (9\,pc at 8\,kpc) to be 250\,eV/cm$^3$. Given the abundance of early
type stars in the cluster, the effective temperature of the radiation field is $\sim$ 35000 K~\citep{2001A&A...377..631R,2009MNRAS.399.1175C}.

The infrared energy density is likely dominated by the emission from
G0.18$-$0.04, the so$-$called ``sickle''. Its luminosity is
$\sim~3.5~\times 10^{40}$~erg/s with a temperature $T_{\rm FIR}$ (FIR:
far infrared) of about 50\,K~\citep{1997A&A...317..441P}.  At the
projected distance of the PWN candidate this implies an energy density
of 50~eV/cm$^3$. This is to be considered a lower limit of the actual
density at the PWN position since the rest of the bubble should
contribute as well to the FIR emission. In the following, we consider an FIR radiation
density of 100 eV/cm$^3$, consistent with values estimated for the CMZ
at large~\citep{2012MNRAS.423.3512C}.  With such large radiation
densities, the evolution of the nebula is likely driven by inverse
Compton losses which can explain the hard X$-$ray spectrum observed by
\chandra~\citep{2007ApJ...657..302H}.

\begin{figure}
\begin{center}
\includegraphics[width=0.49\textwidth]{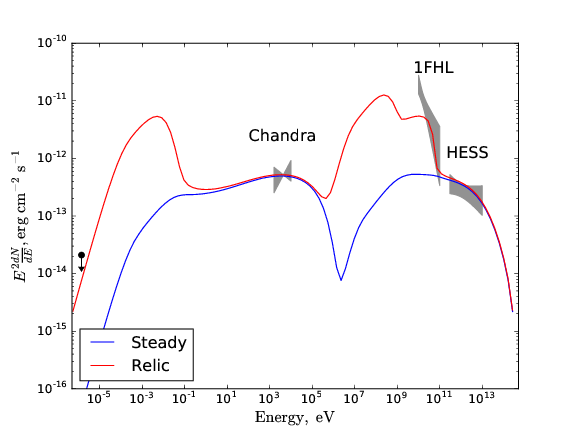}
\caption{Spectral energy distribution of the PWN \arcPWN. The butterflies show the spectral confidence regions of \arcPWN\ measured
  with \chandra~\citep{2002ApJ...581.1148W}, of 1FHL~J1746.3$-$2851~\citep{2013ApJS..209...34A} and of \arcsource.
  The upper limit is obtained from the 90~cm map of \citet{2000AJ....119..207L}. The lines show two models of PWN emission
  for \arcPWN, assuming an inverse Compton dominated evolution of the particles (see
  Sect. \ref{sec:arc_source_nature} for details). The X$-$ray and VHE \gray\ spectra are well
  accounted for by an electron population injected at a few 10$^{35}$~erg/s (steady state model, blue). Part of the flux observed above 10~GeV
  could be explained assuming a significant decay of the pulsar spin down power (relic model, red).} 
\label{fig:arcsource_SED}
\end{center}
\end{figure}

We compute the spectrum radiated by electrons injected by the putative
pulsar as a function of time taking into account pulsar braking \citep[see e.g.][]{2017arXiv170208280H}
and energy dependent losses~\citep{2007ApJ...657..302H}. We use the GAMERA package to
compute the time evolution of the electron
population~\citep{gamera}. The spectrum at injection is chosen to be
an $E^{-2}$ power law extending up to 100~TeV and the injected power is
assumed to be equal to the pulsar spin$-$down power at any time.  The
magnetic field in the nebula is assumed to be constant and we use the
two radiation fields discussed above as well as the cosmic microwave
background.  We first consider the case of a steady injection, and
find that a spin$-$down power of $\dot{E} = 2 \times 10^{35}$~erg/s and
an ambient magnetic field of $B=45\ \mu$G reproduce the X$-$ray and
VHE data (see Fig.~\ref{fig:arcsource_SED}).  If the pulsar has experienced
significant spin down, we might expect a strong flux below 100~GeV
because of the accumulation of cooled electrons, which might explain at
least part of the flux of the \fermi\ source. Assuming dipolar evolution
of the pulsar $\dot{E}$~\citep{2017arXiv170208280H}, we find that a large initial
spin$-$down power of $\dot{E}_0 = 7\times 10^{38}$~erg/s with a
spin$-$down time of $\tau_0 = 500$~years and a pulsar age of 30~kyr can
account for the slope above 10~GeV while preserving an
$\dot{E} \sim 10^{35}$~erg/s required to explain the X$-$ray and VHE \gray\
luminosities (see Fig.~\ref{fig:arcsource_SED}).  Accounting for most
of the GeV flux is more complex and requires extremely high
$\dot{E}_0$.

We note that there are probably several other possible origins for
3FGL~J1746.3$-$2851c, for example GC ridge emission, emission from the
radio arc, or pulsed GeV emission from the putative pulsar. The latter
explanation would require a \gray\ efficiency of the pulsar close to
100\%, if its current $\dot{E}$ is of the order of a few
$10^{35}$~erg/s. This is possible if the actual spin$-$down power
is larger and the fraction of power transferred to the relativistic
wind is not very large.  Finally, we note that
\citet{2013MNRAS.434.1339H} have found an enhancement of soft thermal
X$-$rays around \arcPWN\ that they attribute to the remnant of the
progenitor supernova. This SNR could also contribute to the \gray\
flux. The point$-$like nature of \arcsource\ is difficult to reconcile with
that scenario though.

%%% figure SED
%%% enf figure SED

%%%%%%%%%%%%%%%%%%%%%%%%%%%%%%%%%%%%%%%%%%%%%%%%%%%%%%%%%%%%%%%%%%%%%%%%%%%%%%%%%

%%%%%%%%%%%%%%%%%%%%%%%%%%%%%%%%%%%%%%%%%%%%%%%%%%%%%%%%%%%%%%%%%%%%%%%%%%%%%%%%

\section{Conclusions}

The complete \hess\,I dataset of the GC region provides a very
detailed view of the VHE \gray\ emission in the central 200\,pc of the
Galaxy.  After subtraction of the two main point sources, \GCsource\
and \GPWN, the GC ridge emission appears very clearly distributed like
dense gas as traced by the CS molecule over a projected distance of
140\,pc and fades beyond that. Emission from the main molecular
complexes Sgr~A, B and C is clearly visible as well as fainter
emission from the Sgr~D region. Moreover, the longitude profile now
clearly exhibits a peak at the GC position not visible in the gas 
distribution. The \gray\ emission is a clear sign of a source of CRs
at or near the GC. 

We have built an empirical model of the GC ridge diffuse VHE \gray\ emission
morphology based on an iterative 2D maximum likelihood method.
Starting from a minimal model consisting of the two main point sources
and a diffuse component constructed from a map of dense gas (traced by
the CS molecule) and multiplied by a Gaussian to account for the lower
brightness beyond longitudes of 1$^\circ$, we added Gaussian
components until further components were not found to yield a
significant improvement of the model fit to the data.

Beyond the main diffuse component, which represents 50\% of the total flux, we find
that a large scale emission extending to high latitudes and that does not correlate
with dense gas tracers is required to reproduce the observed morphology.
This component, which accounts for about 30\% of the total flux, is likely the result of CR interactions
with gas in a diffuse phase that dense molecular clouds tracers are not able to map. 
Additional contributions from unresolved sources as well as IC emission from VHE electrons in the region 
can not be excluded. 
An additional component is found for the central 30\,pc. Its flux is 15\% of the total ridge emission.

These three components provide a good model of the diffuse
emission in the central 200\,pc and are consistent with the radial 1/r
gradient of CRs in the CMZ, possibly accelerated at the supermassive 
black hole itself, as concluded in \papIIt. %\citet{GCPeVatron}.
The spectrum extracted in the whole region is well fitted by a power law of photon index 2.3, 
with no significant evidence of a cutoff. This \gray\ spectrum proves the
presence of a PeV$-$scale proton population not only in the central 40 pc, 
but over the entire CMZ including its external more diffuse phase. 

Finally, we detect the new pointlike source \arcsource\ spatially
coincident with the \fermi\ arc source 3FGL~J1746.3$-$2851c and
the X$-$ray PWN candidate \arcPWN\ at the centre of the radio arc IR
bubble, a feature possibly linked to the young and massive Quintuplet
cluster. The intrinsic spectral index of \arcsource\ is found to be $\sim 2.2$
and its luminosity is $7\times 10^{33}$~erg/s at the GC distance, twice
larger than that of the X$-$ray nebula.  This supports a physical
association with the PWN. We have shown that, taking into account the
radiation energy densities in the vicinity of the Quintuplet cluster,
the X$-$ray and VHE \gray\ spectra can be well reproduced by a steady
injection of electrons by a pulsar with a spin down power of a few
times 10$^{35}$~erg/s. It is difficult to explain the GeV emission
in this scenario.  The \fermi\ source might be connected to another
possibly related phenomenon, e.g. the pulsar itself or the underlying
SNR.  \chandra\ has resolved more than a dozen such PWN candidates in
the GC. Albeit fainter than \arcPWN\ they might have detectable
emission in the VHE domain and may be resolved in the future.

The approach presented here does not allow to separate the spectra of the
various components extracted in the morphological analysis, which is a
limitation to understand their origin. 
Due to its enhanced sensitivity and superior angular resolution, the
Cherenkov Telescope Array~\citep[CTA,][]{2013APh....43....3A} will be
crucial to understand the properties of the ridge diffuse emission, as it
will resolve the substructures in much more detail and will be able to
perform combined spectral and morphological analysis in a 3D fit, in 
order to characterise CR acceleration and propagation in the GC region.

%-------------------------------------------------------------------

\begin{acknowledgements}
%\acknowledgments 
The support of the Namibian authorities and of the University of
Namibia in facilitating the construction and operation of H.E.S.S. is
gratefully acknowledged, as is the support by the German Ministry for
Education and Research (BMBF), the Max Planck Society, the German
Research Foundation (DFG), the French Ministry for Research, the
CNRS-IN2P3 and the Astroparticle Interdisciplinary Programme of the
CNRS, the U.K. Science and Technology Facilities Council (STFC), the
IPNP of the Charles University, the Czech Science Foundation, the
Polish Ministry of Science and Higher Education, the South African
Department of Science and Technology and National Research Foundation,
the University of Namibia, the Innsbruck University, the Austrian
Science Fund (FWF), and the Austrian Federal Ministry for Science,
Research and Economy, and by the University of Adelaide and the
Australian Research Council. We appreciate the excellent work of the
technical support staff in Berlin, Durham, Hamburg, Heidelberg,
Palaiseau, Paris, Saclay, and in Namibia in the construction and
operation of the equipment. This work benefited from services provided
by the H.E.S.S. Virtual Organisation, supported by the national
resource providers of the EGI Federation. 
\end{acknowledgements}

%-------------------------------------------------------------------

\bibliographystyle{aa}
\bibliography{GC_ridge_paper_repo}

%\Online

\begin{appendix}
\section{Appendix} %First online appendix

\begin{sidewaystable*}
%\begin{table*}
\caption{The results of the iterative fit are summarised here. 
The definition and optimization of the model of the \gray\ sky describing the data is done through an iterative (n-step)
process,  starting with model $0$ including the background map normalised to 1, the unresolved large scale Galatic emission contribution, and
the two point sources \GCsource\ and \GPWN\, then adding at each step of the fit a new component to improve
the modelization of the emission. At each step, a significance map of the residual and a TS map is created, allowing to 
search for features that should be included in the model. 
  Details of the more relevant parameter values are given for 4 steps of the fitting process.
  The parameters'  68\% confidence intervals are also indicated. The significance of the component detections are indicated
  by the value of  $\Delta$TS /d.o.f. and converted into equivalent significances. The CSTAT/d.o.f. goodness-of-fit is given as an indication
  of the fit quality improvement at each step of the iterative procedure.} 

  \centering
\begin{tabular}{c c c c c c c}      
\hline\hline              
  step &    Model       &       Position                  & Extension         &     Flux                                     &    $\Delta$TS /d.o.f.& CSTAT   \\\\
       &   components   &      (Galactic Coordinates)     & (Degrees)         &    $ (10^{-12} \rm cm^{-2} \rm s^{-1} \rm  TeV^{-1} )$  &     significance     &         \\
          \hline
          \hline
0      &   \GPWN        &  l $= 0.86^{\circ} \pm 0.005^{\circ}_{stat}   $   &                                             & $1.04  \pm 0.4_{stat}$    &           &  1.13  \\
       &                &  b $=0.07^{\circ} \pm 0.003^{\circ}_{stat}   $   &        &                                       &           &        \\
       & \GCsource      &  l $=  359.94^{\circ} \pm 0.002^{\circ}_{stat}$   &                                          &  $3.81 \pm 0.4_{stat}$   &           &        \\
       &                &  b $= -0.05^{\circ}  \pm 0.003^{\circ}_{stat} $   &        &                                       &           &        \\
\hline 
\hline 
1     &  \GPWN          & l $= 0.86^{\circ} $,  b$=0.07^{\circ} $                              &                          & $0.89  \pm 0.04_{stat}$  & 1640/2  &   1.043   \\
      &  \GCsource      & l $=  359.94^{\circ}  $,  b$= -0.05^{\circ} $                         &                        &  $3.1 \pm 0.4_{stat}$ &   40.5 $\sigma$  &   \\
      & DGC %CS\,$\times$\,Gauss 
                                  &     & $\sigma  = 0.77^{\circ}\pm0.2^{\circ}_{stat} $                                       & $4.9  \pm 1.0_{stat}$ & &\\
\hline
\hline
2     &  \GPWN          & l $= 0.86^{\circ} $, b$=0.07^{\circ} $                                &                           & $0.89  \pm 0.04_{stat}$  &  81.8/2  &  1.04 \\
      & \GCsource       & l $= 359.94^{\circ}  $,  b$= -0.05^{\circ} $                           &                         &  $2.92 \pm 0.4_{stat} $ & 8.7  $\sigma$   &  \\
      & DGC             &     & $\sigma  = 0.98^{\circ}\pm0.18^{\circ}_{stat}$                                            & $5.0  \pm 1.0_{stat} $ &  &\\
      &  CC             &     & $\sigma  = 0.12^{\circ}\pm0.01^{\circ}_{stat}$                                             & $1.12 \pm 0.5_{stat} $ & &\\
\hline
\hline
3   & \GPWN           & l $= 0.86^{\circ} $, b$=0.069^{\circ} $                                                  &            & $0.87  \pm 0.04_{stat}$  &  33/3   & 1.039   \\
      & \GCsource    & l $=  359.94^{\circ}  $, b$= -0.05^{\circ} $                                           &            & $2.9 \pm 0.4_{stat}$        &  5.1 $\sigma$ &    \\
      & DGC              &     & $\sigma  = 0.98^{\circ}\pm0.17^{\circ}_{stat}$                                            & $4.4  \pm 1.0_{stat}$       & &\\
      & CC                &     & $\sigma  = 0.12^{\circ}\pm0.01^{\circ}_{stat}$                                             & $1.1 \pm 0.05_{stat}$      & &\\
      & LSC              &     &  $\sigma_{\rm x} = {{1.09^{\circ}}^{+0.02^{\circ}}_{-0.01^{\circ}}}_{stat}$     & $2.68 \pm 0.6_{stat}$      & & \\
      &                     &     &  $\sigma_{\rm y} = 0.25^{\circ}\pm0.05^{\circ}_{stat}$  & & &\\ 
\hline
\hline
4     & \GPWN           & l $= 0.86^{\circ} $, b$=0.069^{\circ} $                                 &                            & $0.88  \pm 0.04_{stat}$  &  48.5/5   & 1.039    \\
      & \GCsource       & l $=  359.94^{\circ}  $, b$= -0.05^{\circ} $                            &                         & $2.9 \pm 0.4_{stat}$       &  5.9 $\sigma$ &    \\
      & DGC             &     & $\sigma  = 1.11^{\circ}\pm0.17^{\circ}_{stat}$                                              & $4.3  \pm 0.9_{stat}$ & &\\
      & CC              &     & $\sigma  = 0.11^{\circ}\pm0.01^{\circ}_{stat}$                                               & $1.03 \pm 0.05_{stat}$ & &\\
      & LSC              &     &  $\sigma_{\rm x} = {{0.97^{\circ}}^{+0.04^{\circ}}_{-0.02^{\circ}}}_{stat}$      & $2.68 \pm 0.6_{stat}$ & & \\
      &                 &     &  $\sigma_{\rm y} = 0.22^{\circ}\pm0.06^{\circ}_{stat}$                                     & & &\\
      & \arcsource      & l $= 0.14^{\circ}\pm 0.01^{\circ}_{stat}$   & $\sigma_{\rm x} = 0.03^{\circ} \pm 0.03^{\circ}_{stat}$ & $0.24 \pm 0.03_{stat} $ & &\\
      &                 & b $= -0.11^{\circ}\pm 0.02^{\circ}_{stat}$  & $\sigma_{\rm y} = 0.02^{\circ}\pm 0.02^{\circ}_{stat}$ &  & &\\
\hline\hline
\end{tabular}
\label{table:2}   
%\end{table*}
\end{sidewaystable*}

\end{appendix}

\end{document}